\documentclass{article}
\usepackage{arxiv}
\usepackage[utf8]{inputenc}
\usepackage[T1]{fontenc}
\usepackage[hidelinks]{hyperref}
\usepackage{url}
\usepackage{booktabs}
\usepackage{amsfonts}
\usepackage{amsmath}
\usepackage{nicefrac}
\usepackage{microtype}
\usepackage{cleveref}
\usepackage{graphicx}
\usepackage{natbib}
\usepackage{doi}
\usepackage{wrapfig}
\usepackage{xcolor}
\usepackage[final]{changes}

\title{\replaced{Automated Detection of Ripple-Like Mesospheric Wave Structures in OH Airglow Images Using a Squeeze-and-Excitation Convolutional Neural Network}{Automated Detection and Climatological Analysis of Ripple-Scale Gravity Wave Instabilities Using a Squeeze-and-Excitation Convolutional Neural Network}}

\author{
Jiahui~Hu$^{1}$, 
Alan~Liu$^{1}$, 
Adriana~Feener$^{1}$, 
Wenjun~Dong$^{1,*}$ \\
$^{1}$Embry-Riddle Aeronautical University, Daytona Beach, FL, USA \\
$^{*}$Corresponding author: \texttt{Wenjun Dong, DONGW1@erau.edu}
}

\renewcommand{\shorttitle}{\replaced{Automated Ripple-Like Structure Detection}{TEXT}}

\begin{document}
\maketitle
\begin{abstract}
\replaced{All-sky OH airglow imaging provides two-dimensional observations of mesospheric wave morphology near the nominal OH emission peak altitude. Small, localized ripple-like structures are difficult to catalogue consistently through manual inspection and may represent either instability-related features or secondary gravity waves. We present an automated detection framework based on a squeeze-and-excitation convolutional neural network (SE-CNN) trained on 41 × 41-pixel patches extracted from 512 × 512-pixel all-sky images acquired at Yucca Ridge Field Station. Signed differences between consecutive images separated by approximately 2 min are normalized using spatial median-absolute-deviation scaling before classification. The trained model is applied to complete images through sliding-window inference, followed by score-map reconstruction, component filtering, and spatial matching. On the held-out patch test set, the classifier achieved an accuracy of 0.92, precision of 0.89, recall of 0.93, and F1 score of 0.92. Within a selected cohort of manually positive validation frames, 299 of 415 evaluable frames contained at least one reference–automated spatial overlap, and 324 of 559 eligible reference supports received a one-to-one automated-component assignment. In a separate blinded single-reviewer audit of disagreement patches, 411 of 543 model-candidate/reference-unmarked crops were classified as clear ripple-like, corresponding to a clear-ripple fraction of 75.7\% with a night-cluster bootstrap 95\% interval of 70.4\%–80.2\%. Because the audit was restricted to selected disagreement cases from manually positive frames, this value is not an archive-wide precision estimate. The framework applies a learned morphological criterion consistently across analyzed images while remaining dependent on the manually labeled training data. The automated detections identify ripple-like image morphology and do not determine the underlying physical generation mechanism.}{All-sky OH airglow imaging provides two-dimensional observations of mesospheric gravity wave structure near 87 km altitude. Ripple-scale instability signatures, characterized by 5–15 km horizontal wavelengths and short lifetimes, are particularly difficult to identify consistently using manual inspection. In this study, we develop a reproducible, automated detection framework based on a squeeze-and-excitation convolutional neural network (SE-CNN) trained on 41 x 41 pixel image patches, to identify ripple-scale structures in 512 × 512 pixel all-sky airglow images acquired at Yucca Ridge Field Station (40.7◦N,104.9◦W). The time-differenced images are normalized using a robust median-absolute-deviation (MAD) scaling procedure to mitigate star contamination and background variability. The model is trained and validated on manually annotated ripple and non-ripple patches, then evaluated using independent test subsets. The automated detection is performed using a sliding-window approach with spatial and temporal clustering criteria for event definition. At the patch level, the classifier achieves 92\% F1-score with high precision and recall. At the event level, automated detections recover approximately 90\% of manually identified ripple events while identifying additional low-amplitude occurrences. Validated against previous manual identification study, the automated detection catalog enables objective quantification of ripple occurrence frequency, seasonal modulation, and lifetime distributions. By emphasizing methodological transparency, calibration considerations, and validation metrics, this framework establishes a scalable measurement technique for systematic detection of mesospheric instability signatures in long-term airglow image archives.}
\end{abstract}


\section{Introduction}

\deleted{All-sky airglow imaging has become a primary observational technique for investigating mesospheric gravity waves and instability processes. By capturing spatial variations in OH nightglow emission near ~87 km altitude, fish-eye imagers provide two-dimensional maps of horizontal wave structure with high temporal cadence (typically 1–2 min). These measurements enable direct visualization of wave fronts, instability signatures, and short-lived dynamical features that are otherwise inaccessible through vertically profiling instruments alone. Theoretical and observational reviews of gravity wave dynamics emphasize that convective and shear instabilities arise when wave amplitudes become sufficiently large, producing localized overturning and turbulence structures detectable in airglow imagery (\cite{fritts2003gravity}).}
\added{All-sky airglow imaging is widely used to investigate mesospheric gravity waves and related dynamical processes. By capturing spatial variations in OH nightglow emission near the nominal emission peak around 87~km, fish-eye imagers provide two-dimensional maps of horizontal wave morphology at a cadence of typically 1--2~min. Theoretical and observational studies indicate that sufficiently large gravity-wave amplitudes can produce convective and shear instabilities, localized overturning, and turbulence \citep{fritts2003gravity}. Ripple-like structures in OH airglow images have therefore traditionally been interpreted as signatures of instability and gravity-wave breaking \citep{taylor1990origin,yamada2001breaking,li2005concurrent}. However, similar small-scale patterns may also represent secondary gravity waves \citep{li2017characteristics}. Because these mechanisms can produce comparable horizontal morphology, airglow images alone generally do not permit an unambiguous distinction between instability features and secondary gravity waves. We therefore use ``ripple-like structure'' as a morphological term throughout this study. Their small image-plane scale makes identification sensitive to normalization, star contamination, cloud interference, and geometric distortion.}

\deleted{Previous statistical studies of ripple occurrence have relied on manually curated event identification within multi-year image archives (\cite{li2017characteristics}). \citet{yue2010seasonal} analyzed multi-year OH airglow observations over Colorado and Japan and found that ripple occurrence maximizes near solstices and minimizes near equinoxes, consistent with seasonal modulation of gravity wave activity. While these analyses established important seasonal and morphological characteristics, manual classification limits scalability and introduces observer-dependent variability. In addition, faint or short-lived ripple signatures may be systematically underdetected, potentially biasing climatological statistics. From a measurement standpoint, an automated and reproducible detection framework is needed to (i) apply a consistent decision criterion across extensive datasets, (ii) reduce subjective bias, and (iii) quantify detection performance using objective metrics.}
\added{Previous statistical studies have relied on manually curated ripple catalogs from multi-year image archives \citep{li2017characteristics}. For example, \citet{yue2010seasonal} reported seasonal variability in OH airglow observations over Colorado and Japan. Such catalogs are physically informed, but they depend on the annotation criteria, experience, and decisions of the observer, making exact reproduction across large datasets difficult. A classifier trained on these annotations does not eliminate the subjectivity or uncertainty of the original labels. Its methodological advantage is instead that it applies the learned decision criterion consistently across the analyzed images. Accordingly, the present study evaluates agreement with a manual reference catalog using predefined patch-, frame-, object-, and disagreement-patch-level measures.}

\deleted{Recent applications of machine learning to atmospheric imaging have demonstrated automated extraction of large-scale gravity wave patterns from airglow and satellite observations \cite{lai2019automatic, xiao2024atmospheric, okui2025cnn}. These approaches, however, primarily target coherent, high-contrast wave fronts extending over tens to hundreds of kilometers. Ripple-scale instability signatures differ fundamentally: they are low-amplitude, spatially localized, and embedded within heterogeneous backgrounds. Their detection requires enhanced sensitivity to weak, quasi-periodic banded modulations while maintaining robustness to instrumental artifacts and background variability. A recent study by \cite{prasad2025ripple} applied the YOLOv8 object detection model to shortwave infrared OH airglow images over Mt. Abu to automatically identify ripple structures, achieving a mean average precision (mAP) of 0.62 and demonstrating the potential of transfer learning–based object detectors for mesospheric ripple localization. Recent work on ASI cloud segmentation using enhancement fully convolutional networks demonstrates that deep learning–based approaches provide improved robustness over traditional threshold methods, particularly under nocturnal low-signal conditions \cite{shi2019diurnal}.}
\added{Machine-learning methods have increasingly been applied to wave extraction in airglow imagery. \citet{lai2019automatic} developed a machine-learning approach for extracting gravity-wave patterns from all-sky images, while \citet{prasad2025ripple} used a YOLOv8 object detector to localize ripple structures in shortwave-infrared OH airglow images. \citet{wust2026extraction} used a modified YOLOv7 framework to identify spatially confined small-scale wave-like structures and estimate their spatial properties in high-resolution airglow images. These studies differ in instrument characteristics, spatial resolution, label definitions, and output units. The present method instead classifies small image patches and reconstructs a full-image score field through sliding-window inference.}

\deleted{From a signal-processing perspective, standard convolutional neural networks (CNNs) treat all feature channels equally, which may limit discrimination of subtle instability signatures relative to dominant large-scale structures. To address this limitation, we adopt a squeeze-and-excitation (SE) mechanism (\cite{hu2018squeeze}) that introduces adaptive channel-wise recalibration. By incorporating global context into feature weighting, the SE block enhances channels that capture coherent ripple morphology while suppressing noise-dominated responses. This architecture is particularly suited to ripple detection, where physically meaningful features occupy a narrow spatial scale range and exhibit low contrast.
In this study, we develop a measurement-driven, SE-enhanced CNN framework for automated ripple detection in all-sky OH airglow imagery obtained at Yucca Ridge Field Station (YRFS). The dataset consists of 512 × 512 pixel fish-eye images acquired at 2 min cadence, using a CCD system binned from 1024 × 1024 resolution and equipped with a ~180° field-of-view lens. Images are normalized using a robust Median-Absolute-Deviation (MAD) scaling approach to mitigate star contamination and background variability. The model is trained on manually annotated ripple and non-ripple patches and applied to full-image sliding-window detection with explicit clustering criteria for spatial and temporal event definition.
In this study, we develop a measurement-driven, SE-enhanced CNN framework for automated ripple detection in all-sky OH airglow imagery obtained at Yucca Ridge Field Station (YRFS). The dataset consists of 512 × 512 pixel fish-eye images acquired at 2 min cadence, using a CCD system binned from 1024 × 1024 resolution and equipped with a ~180° field-of-view lens. Images are normalized using a robust Median-Absolute-Deviation (MAD) scaling approach to mitigate star contamination and background variability. The model is trained on manually annotated ripple and non-ripple patches and applied to full-image sliding-window detection with explicit clustering criteria for spatial and temporal event definition.}
\added{Conventional CNNs learn multiple feature channels but do not necessarily perform explicit channel-wise recalibration using global contextual information. We therefore adopt a squeeze-and-excitation (SE) mechanism \citep{hu2018squeeze}, which derives channel-wise weights from globally aggregated feature responses. This mechanism allows the network to adaptively reweight feature channels according to their learned contributions to the classification task.}

\deleted{Detection performance is evaluated using independent training, validation, and test subsets, and results are compared at both patch and event levels with a previously published manual ripple catalog (Li et al., 2017). The resulting automated event catalog enables objective quantification of ripple occurrence frequency, seasonal modulation, lifetime characteristics, and occurrence time dependence. More broadly, this framework establishes a scalable technique for systematic identification of instability signatures in airglow imagery and provides a reproducible foundation for constructing long-term climatologies of mesospheric gravity wave breaking activity.}
\added{In this study, we develop an SE-CNN framework for detecting ripple-like morphology in all-sky OH airglow images acquired at Yucca Ridge Field Station (YRFS). Signed two-minute image differences are normalized using median-absolute-deviation (MAD) scaling, classified in overlapping $41\times41$-pixel windows, and converted into retained image-plane components. Performance is evaluated first on a held-out patch test set and then, within a selected cohort of manually positive validation frames, through frame-level spatial recovery, maximum-cardinality one-to-one object matching, and a blinded single-reviewer audit of disagreement patches. Provisionally linked tracks are summarized descriptively, but the selected cohort does not provide archive-wide occurrence rates or an exposure-normalized climatology. The purpose of the study is to establish a reproducible image-analysis method rather than to determine the physical generation mechanism of individual structures.}

\section{Methodology}
\subsection{Image preprocessing and patch generation}

\deleted{The airglow images used in this study were obtained from the all-sky imager located at YRFS in Fort Collins, Colorado ($40.7^\circ N, 104.9^\circ W$) (\cite{nakamura2005simultaneous}). The imager captures the OH airglow emission every 2 minutes, with a peaking altitude at approximately 87~km (\cite{taylor1991near}). It consists of a Nikon fish-eye Nikkor ($f=8mm, f/2.8$) lens (roughly 180$^\circ$ field-of-view) and a CCD camera with a Kodak $1024\times1024$ CCD chip binned to $512\times512$.}
\added{The OH airglow images were obtained with the all-sky imager at YRFS near Fort Collins, Colorado ($40.7^\circ$N, $104.9^\circ$W) \citep{nakamura2005simultaneous}. The nominal image cadence was 2~min. OH emission typically peaks near 87~km, although the layer altitude varies temporally and geographically \citep{wust2020variability,garcia2017mesospheric,von2015lunar,baker1988rocket}. The instrument used a Nikon 8~mm $f/2.8$ fish-eye lens with an approximately $180^\circ$ field of view and a Kodak $1024\times1024$-pixel CCD binned to $512\times512$ pixels. The complete fish-eye field nominally approaches a zenith angle of $90^\circ$, whereas the present validation was restricted to a centered 125-pixel-radius region of interest. Because of the fish-eye projection, the horizontal distance represented by one pixel and the physical area represented by a $41\times41$-pixel patch vary with zenith angle.}

\deleted{To prepare the train/test datasets, each intensity frame is normalized and clipped into the range of $[0,1]$:
}
\added{For each target image $I_t(x,y)$, the immediately preceding valid image $I_{t-\Delta t}(x,y)$, acquired approximately 2~min earlier, was identified. If no valid predecessor was available, the target image was excluded. A signed time-difference image was then calculated as}

\begin{equation}
    D_t(x,y)=I_t(x,y)-I_{t-\Delta t}(x,y),
    \label{eq:difference}
\end{equation}

Time differencing acts as a temporal high-pass operation whose response depends on the approximately 2~min image cadence and the apparent phase speed of the observed structures. Stationary and slowly varying features are
suppressed, whereas structures that change in intensity or translate between frames are enhanced. Slowly evolving ripple-like structures may therefore be
attenuated, while translating wave fronts may produce paired positive and negative signatures at the locations entered and vacated between consecutive frames. Rapid motion may also reduce the spatial coherence of the residual pattern. Thus, the differencing step modifies the effective selection function of the detector and preferentially retains structures that produce sufficient frame-to-frame change.

The median and median absolute deviation (MAD) were then calculated 
spatially over all pixels in the complete time-difference image:
\begin{equation}
\operatorname{MAD}(D_t)
=
\operatorname{median}_{x,y}
\left(
\left|
D_t(x,y)
-
\operatorname{median}_{x,y}\!\left(D_t\right)
\right|
\right).
\label{eq:mad}
\end{equation}

The time-difference image was then normalized and clipped to the interval
$[0,1]$ according to
\begin{equation}
D_{\mathrm{norm}}(x,y)
=
\operatorname{clip}
\left(
\frac{
D_t(x,y)
-
\operatorname{median}_{x,y}\!\left(D_t\right)
}{
8\,\operatorname{MAD}(D_t)
}
+ 0.5,\,
0,\,
1
\right).
\label{eq:normalization}
\end{equation}

\deleted{The MAD provides a robust measure of intensity variability that is less sensitive to localized bright features (e.g., stars or instrumental artifacts) than the standard deviation. The factor of 8 was empirically selected to scale typical background variability into the $[0,1]$ range while preserving ripple-scale contrast. Clipping suppresses extreme intensity excursions and improves numerical stability during training.
Each normalized all-sky image is divided into a set of small overlapping patches. Each patch is $41\times 41$ pixels in size, which corresponds to a roughly fixed km-scale area. Patches were centered either at known ripple locations from manual annotations of training images, or at randomly sampled background locations with no visible ripple, ensuring a balanced representation of ripple and non-ripple examples in the training data. Each patch is labeled as ``ripple'' or ``non-ripple''. The amounts of ripple and none-ripple patch are approximately equivalent, around 1500 for both category. To enhance generalization, random horizontal and vertical flips and small rotations ($\pm 10^\circ$) were applied during training.
Each normalized all-sky image is divided into a set of small overlapping patches. Each patch is $41\times 41$ pixels in size, which corresponds to a roughly fixed km-scale area. Patches were centered either at known ripple locations from manual annotations of training images, or at randomly sampled background locations with no visible ripple, ensuring a balanced representation of ripple and non-ripple examples in the training data. Each patch is labeled as ``ripple'' or ``non-ripple''. The amounts of ripple and none-ripple patch are approximately equivalent, around 1500 for both category. To enhance generalization, random horizontal and vertical flips and small rotations ($\pm 10^\circ$) were applied during training.}
\added{The MAD is less sensitive than the standard deviation to localized bright outliers such as stars and instrumental artifacts. The factor of 8 was selected empirically to map typical background variability into the interval $[0,1]$ while retaining weak banded contrast. The same complete-frame normalization was applied during patch generation and full-image inference.}

\added{The reconstructed patch dataset contained 725 positive examples, with one $41\times41$-pixel patch centered on each available corrected reference-annotation coordinate. For each positive patch, one background patch was selected from the same source image using a seeded deterministic sampling procedure. Background centers were restricted to the central field of view, were required to be at least 60 pixels from every reference annotation in the corresponding image, and were separated from other selected background centers. The background patches were not independently adjudicated and may therefore contain unannotated morphology or artifacts. Consequently, the held-out patch metrics characterize discrimination under this specific reference-centered positive and sampled-background procedure rather than against an independently confirmed hard-negative set.}

\added{Horizontal and vertical flips and rotations of up to $\pm10^\circ$ were applied only to the training partition. Because the fish-eye projection does not provide a constant physical scale across the image, the classifier operates in image coordinates rather than assuming that every patch represents the same horizontal area.}

\subsection{\replaced{Neural network architecture, training, and validation strategy}{Neural Network Architecture, Training and Validation strategy}}
\subsubsection{SE-CNN architecture}
\begin{figure}[!t]
    \centering
    \includegraphics[width=0.8\linewidth]{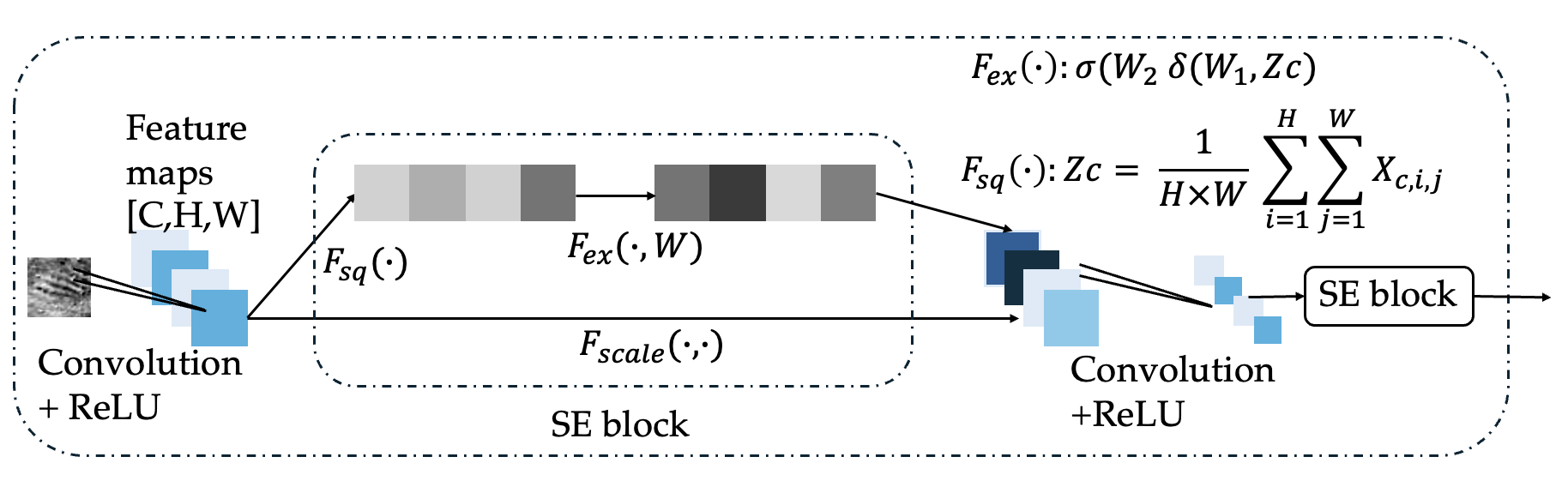}
    \caption{
    Schematic illustration of the squeeze-and-excitation (SE) block used in the convolutional neural network. Input feature maps of size $[C, H, W]$ are first passed through a convolutional layer followed by a ReLU activation. The squeeze operation $F_{sq}(\cdot)$ applies global average pooling over the spatial dimensions to generate channel-wise descriptors $Z_c$. These descriptors are then processed by the excitation operation $F_{ex}(\cdot)$, which consists of two fully connected layers with a nonlinear activation and a sigmoid function to produce channel-wise weights. The resulting weights are applied through the scaling operation $F_{scale}(\cdot)$ to recalibrate the original feature maps, which are subsequently passed to the next convolutional layer.
    }
    \label{fig:SE-CNN}
\end{figure}

\deleted{We implemented a CNN classifier augmented with a SE block (Figure~\ref{fig:SE-CNN}). The SE mechanism adaptively recalibrates channel-wise feature responses through global average pooling followed by learned gating, allowing the network to emphasize feature maps that capture coherent banded ripple structures while suppressing noise-dominated or large-scale background features. To objectively and efficiently identify gravity wave breaking signatures in
large airglow image datasets,we have developed a simple CNN-based model trained using hundreds of manually identified ripple structures in a 41×41 pixel image. The CNN consists of six convolutional layers with 16, 32, 32, 64, 64, and 128 filters, respectively, each using 3×3 kernels, stride 1, and same padding. Each convolution is followed by ReLU activation and batch normalization. Max pooling (2×2) is applied after every two convolutional layers. A fully connected layer with sigmoid activation produces the final ripple probability. The convolutional feature maps after each layer are passed through an SE block and then to a fully connected layer with sigmoid activation to produce a ripple probability. 
}
\added{The ripple classifier is a convolutional neural network (CNN) augmented with squeeze-and-excitation (SE) blocks (Figure~\ref{fig:SE-CNN}). A CNN applies trainable convolutional filters to local neighborhoods, with each filter producing a spatial feature map. The collection of feature maps forms the channel dimension. Throughout this paper, $C$, $H$, and $W$ denote the number of channels, feature-map height, and feature-map width, respectively. The network receives one $41\times41$-pixel grayscale patch. Six convolutional layers with $3\times3$ kernels extract progressively more abstract image features. Each convolution is followed by batch normalization and a rectified linear unit (ReLU), $\operatorname{ReLU}(x)=\max(0,x)$. Max pooling is applied after every two convolutional layers. After selected convolutional stages, an SE block learns channel-wise weights from globally aggregated feature responses. The operation changes the relative contribution of the feature channels but does not by itself establish the physical origin of a detected structure.}
\begin{table}[ht]
\centering
\caption{Architecture of the SE-CNN used for ripple classification.}
\label{tab:cnn_architecture}
\begin{tabular}{llll}
\hline
\textbf{Stage} & \textbf{Operation} & \textbf{Output} & \textbf{Description} \\
\hline
Input & --- & $1 \times 41 \times 41$ &
Grayscale image patch \\
\hline
Conv1 & $3\times3$, 16 filters &
$16 \times 41 \times 41$ &
ReLU + BatchNorm \\
Conv2 & $3\times3$, 32 filters &
$32 \times 41 \times 41$ &
ReLU + BatchNorm + SE block \\
MaxPool & $2\times2$ &
$32 \times 20 \times 20$ &
Downsampling \\
\hline
Conv3 & $3\times3$, 32 filters &
$32 \times 20 \times 20$ &
ReLU + BatchNorm \\
Conv4 & $3\times3$, 64 filters &
$64 \times 20 \times 20$ &
ReLU + BatchNorm + SE block \\
MaxPool & $2\times2$ &
$64 \times 10 \times 10$ &
Downsampling \\
\hline
Conv5 & $3\times3$, 64 filters &
$64 \times 10 \times 10$ &
ReLU + BatchNorm \\
Conv6 & $3\times3$, 128 filters &
$128 \times 10 \times 10$ &
ReLU + BatchNorm + SE block \\
\hline
Global Pool & Adaptive average pooling &
$128$ &
Channel aggregation \\
Fully Connected & Linear layer &
2 logits &
Ripple / non-ripple scores \\
Softmax & --- &
\replaced{2 normalized scores}{2 probabilities} &
\replaced{Uncalibrated class scores}{Classification output} \\
\hline
\end{tabular}
\end{table}

The Squeeze-and-Excitation (SE) block is a lightweight architectural unit designed to enhance the representational capacity of convolutional neural networks by explicitly modeling channel-wise feature interdependencies. Given an intermediate feature tensor 
$\mathbf{U} \in \mathbb{R}^{C \times H \times W}$, 
the SE block performs three sequential operations: \emph{squeeze}, \emph{excitation}, and \emph{scaling}.

First, the squeeze operation aggregates global spatial information using global average pooling:
\begin{equation}
z_c = F_{sq}(\mathbf{U})_c 
= \frac{1}{H W} \sum_{i=1}^{H} \sum_{j=1}^{W} U_{c,i,j},
\quad c = 1, \dots, C,
\end{equation}
producing a channel descriptor vector 
$\mathbf{z} = [z_1, \dots, z_C]^\top \in \mathbb{R}^{C}$, 
which captures global contextual information for each channel.

Next, the excitation operation models nonlinear channel dependencies through a two-layer fully connected bottleneck with reduction ratio $r$:
\begin{equation}
\mathbf{s} = F_{ex}(\mathbf{z}) 
= \sigma\!\left( \mathbf{W}_2 \, \delta(\mathbf{W}_1 \mathbf{z}) \right),
\end{equation}
where $\mathbf{W}_1 \in \mathbb{R}^{\frac{C}{r} \times C}$,
$\mathbf{W}_2 \in \mathbb{R}^{C \times \frac{C}{r}}$,
$\delta(\cdot)$ denotes the ReLU activation, and 
$\sigma(\cdot)$ denotes the sigmoid function.
The resulting vector 
$\mathbf{s} = [s_1, \dots, s_C]^\top \in (0,1)^C$
provides adaptive channel-wise weights. Finally, the scaling operation recalibrates the original feature maps by channel-wise multiplication:
\begin{equation}
\tilde{U}_c = F_{scale}(\mathbf{U})_c = s_c \cdot U_c,
\end{equation}
or equivalently,
\begin{equation}
\tilde{\mathbf{U}} = \mathbf{s} \odot \mathbf{U},
\end{equation}
where $\odot$ denotes channel-wise multiplication with broadcasting over spatial dimensions. 
The recalibrated tensor $\tilde{\mathbf{U}}$ emphasizes informative feature channels while suppressing less useful ones, and is subsequently passed to the next layer of the network.

\begin{figure}
    \centering
  \includegraphics[width=\linewidth]{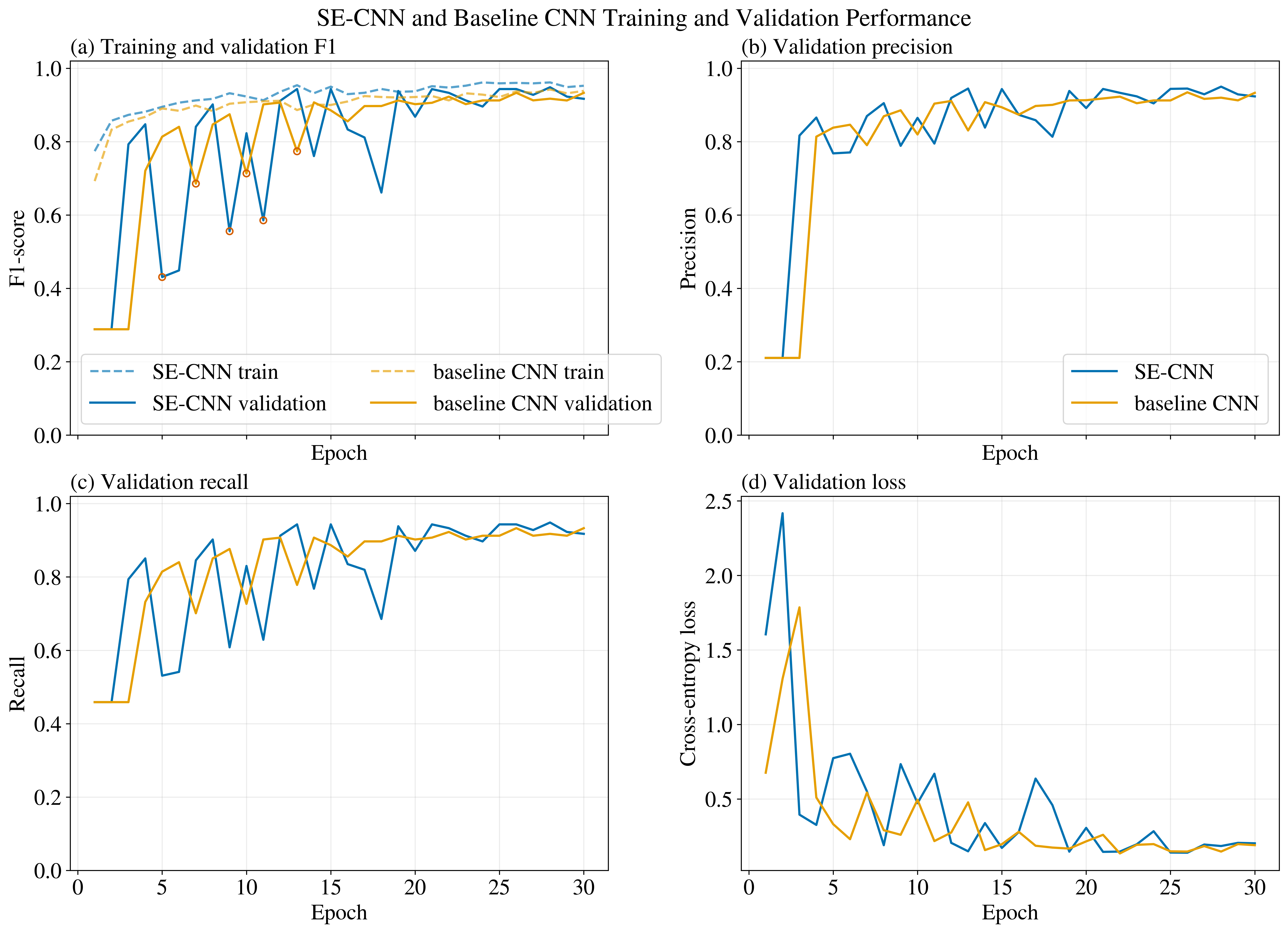}
    \caption{Training history of the SE-CNN and baseline CNN over 30 epochs.
    (a) Training and validation F1 scores, with dashed lines representing training
    performance and solid lines representing validation performance. Open vermillion
    circles indicate epochs at which the validation F1 score decreased by more than
    0.05. (b) Validation precision, (c) validation recall, and (d) validation
    cross-entropy loss. The validation curves were calculated using the saved
    194-patch validation split.}
    \label{fig:cnn-SE-comparison}
\end{figure}

\subsubsection{\replaced{Training and model selection}{Training and Full-image inference}}
\deleted{The model was trained using binary cross-entropy loss and optimized with the Adam algorithm (learning rate $\eta = 10^{-3}$, weight decay $10^{-4}$). Training proceeded for 30 epochs with a batch size of 64. A stepwise learning rate scheduler reduced $\eta$ by a factor of 0.5 every 10 epochs. Early stopping based on validation loss was implemented to prevent overfitting. The dataset was partitioned into independent training (70\%), validation (15\%), and testing (15\%) subsets. Patches derived from the same image were assigned to the same subset to prevent data leakage. All hyperparameter tuning was performed using only the training and validation sets. Model performance on the hold-out test set was evaluated using classification accuracy, precision, recall, and F1-score. These metrics quantify overall correctness, reliability of ripple detections, completeness of ripple recovery, and the harmonic balance between precision and recall, respectively.}
\added{The model was trained using cross-entropy loss and the Adam optimizer with a learning rate of $10^{-3}$ and weight decay of $10^{-4}$. Training proceeded for 30 epochs with a batch size of 64, and the learning rate was reduced by a factor of 0.5 every 10 epochs. Complete observing nights and connected annotation groups were assigned wholly to the training, validation, or test partition to reduce leakage among temporally or spatially related patches. The target split was approximately 70\%/15\%/15\%. The checkpoint with the highest validation F1 score was selected using the validation partition, and the held-out test partition was evaluated once after model selection.}

\added{Patch-level performance was summarized using accuracy, precision, recall, and F1 score. These quantities describe performance under the stated reference-centered positive and sampled-background procedure. They do not estimate archive-wide precision, specificity, or quiet-frame false-alarm rate.}

\subsubsection{\added{Full-image inference, component extraction, and temporal linking}}
\label{subsubsection:full-image-inference}
\deleted{After training, the SE-CNN model was applied to full images using a sliding-window approach with a stride of 5 pixels. Each patch was assigned a ripple probability, and patches exceeding a classification threshold of 0.99 were labeled as ripple-positive. Clusters separated by fewer than 8 pixels ($\approx 8km$) were merged and counted as a single ripple occurrence at that timestamp. To define ripple events in time, spatially overlapping clusters detected in consecutive images were grouped, accounting for ripple motion. A ripple event was therefore defined as a temporally continuous sequence of spatially coherent ripple detections occurring in approximately the same geographic region. This procedure allows automated identification of ripple presence, approximate location, and temporal evolution across the full airglow dataset.}
\added{During full-image inference, $41\times41$-pixel windows were evaluated with a stride of 5 pixels. The positive-class softmax output was used as an uncalibrated model score rather than as a numerical probability.}

\added{Pixels with positive-class model scores greater than 0.99 were thresholded to form candidate regions. Neighboring thresholded detections separated by fewer than 8 pixels were merged before component measurement. Connected components were retained when their pre-clipping area was at least 125 pixels and their eccentricity was no greater than 0.95. Analysis was restricted to the centered 125-pixel-radius region of interest.}

\added{For the provisional temporal analysis, retained components in consecutive frames were linked when their temporal separation was no greater than 121.5~s and their centroid displacement was no greater than 10 pixels. The resulting tracks are products of this linking procedure and are not assumed to be independently verified continuous physical events.}

\subsubsection{\added{Frame- and object-level agreement}}
\added{Each reference annotation was represented by a $41\times41$-pixel support centered on its catalog coordinate, and each retained automated detection was represented by the filled mask of its thresholded component. A reference support and an automated component were considered spatially overlapping when their masks shared at least one pixel in the same frame. No minimum overlap fraction, intersection-over-union threshold, or centroid-distance criterion was imposed.}

\added{At the frame level, an evaluable manually positive frame was classified as spatially recovered when at least one eligible reference support overlapped at least one retained automated component. At the object level, maximum-cardinality one-to-one assignment was performed among all spatially overlapping support--component pairs. Manual-object recall was calculated as the number of assigned reference supports divided by the number of eligible reference supports. The automated-component assignment fraction was reported separately and was not interpreted as precision because the reference catalog was not assumed to be exhaustive.}

\subsubsection{\replaced{Single-reviewer blinded audit of disagreement patches}{Single-assessor blinded audit of disagreement patches}}
\deleted{}
\added{To characterize morphology associated with reference--automated disagreement, we conducted a single-reviewer blinded audit of 720 $41\times41$-pixel patches. The audit contained two prespecified strata. The first comprised 177 reference-marked patches whose annotation support had zero overlap with any final retained automated component. The second comprised 543 retained automated candidates with model scores greater than 0.99 and zero overlap with every reference support. All candidates in the second stratum came from manually positive validation frames. The strata are referred to as reference-marked/component-unmatched and model-candidate/reference-unmarked, respectively.}

\deleted{}
\added{Patches were presented using blind identifiers, and the reviewer assigned one primary morphology label to each case: clear ripple-like, non-ripple or artifact. All 720 expected blind identifiers were completed exactly once. The interface also contained confidence, band, coherence, and artifact-subtype fields. Because every value in those secondary fields remained identical to the interface's automatic preset, they were not treated as independently entered judgments and were excluded from the analysis.}

\deleted{}
\added{Category counts and percentages were calculated separately for the two audit strata. Uncertainty was estimated by resampling observing nights as clusters while retaining all reviewed patches from each sampled night. Because the audit used one reviewer, inter-reviewer agreement and consensus classifications were unavailable. The strict-zero-overlap rule used for audit selection differs from maximum-cardinality one-to-one assignment; therefore, the audit denominators are not expected to equal the one-to-one unmatched-object counts.}

\section{Results}
\deleted{The performance of the SE-CNN and vanilla CNN ripple detection frameworks was compared at patch level, as Figure \ref{fig:cnn-SE-comparison}. SE-CNN detection framework is evaluated and systematically compared with the manually curated ripple catalog of Li et al. (2017). An example of an all-sky image with ripple probability contour is shown in Fig. \ref{fig:cnn-recognition}. At the patch level, the classifier achieved an accuracy of 0.92, precision of 0.89, recall of 0.93, and F1-score of 0.91 on the independent test set, indicating strong discrimination between ripple and non-ripple image regions. The slightly higher recall relative to precision reflects a modest tendency toward enhanced sensitivity to faint ripple structures, which is desirable for climatological completeness.
}
\added{Figure~\ref{fig:cnn-SE-comparison} compares the training and validation histories of the SE-CNN and a baseline CNN that used the same convolutional architecture but omitted the SE blocks. The baseline was used only as an architectural comparator; the SE-CNN was used for full-image inference. On the held-out patch test set, the SE-CNN achieved an accuracy of 0.92, precision of 0.89, recall of 0.93, and F1 score of 0.92. These metrics characterize the stated reference-centered positive and sampled-background procedure and should not be interpreted as archive-wide precision, quiet-frame false-alarm rate, or performance on an independently adjudicated hard-negative set. Figure~\ref{fig:ripple_recognition} provides representative full-image inference examples, while the frequency of ripple-like morphology among selected disagreement cases was quantified separately through the blinded audit.}

\added{The largest temporary decreases in validation F1 in Figure~\ref{fig:cnn-SE-comparison} were accompanied primarily by decreases in validation recall, whereas precision varied less strongly. Because the validation partition contained 194 patches, a small number of cases crossing the classification threshold could produce visible changes in recall and F1.}

\begin{figure}
    \centering
    \includegraphics[width=\linewidth]{fig3_detection_example_revised.png}
    \caption{\replaced{Each row corresponds to one image frame identified by its acquisition timestamp in YYMMDDHHMMSS format. The columns show the raw current image (left), the MAD-normalized signed time-difference image (center), and the detection/reference overlay (right). Blue contours mark the model-score threshold of 0.99, blue stars mark score-weighted centroids of retained automated regions, and bright-yellow stars mark reference-annotation centers. The dashed bright-yellow circle delineates the centered 125-pixel-radius analysis region. Model scores are used for thresholding and are not interpreted as calibrated probabilities.}{ Each row corresponds to one image frame, identified by its acquisition timestamp in YYMMDDHHMMSS format. The columns show the raw image (left), MAD-normalized image (center), and detection overlay (right). Blue contours indicate a model-score threshold of 0.99, blue stars mark the score-weighted centroids of SE-CNN-detected regions, and bright-yellow stars mark manual recognitions. The dashed bright-yellow circle delineates the centered 125-pixel-radius analysis region. Model scores are used for detection and are not interpreted as calibrated probabilities. }}
    \label{fig:ripple_recognition}
\end{figure}

\begin{figure}
    \centering
    \includegraphics[width=\linewidth]{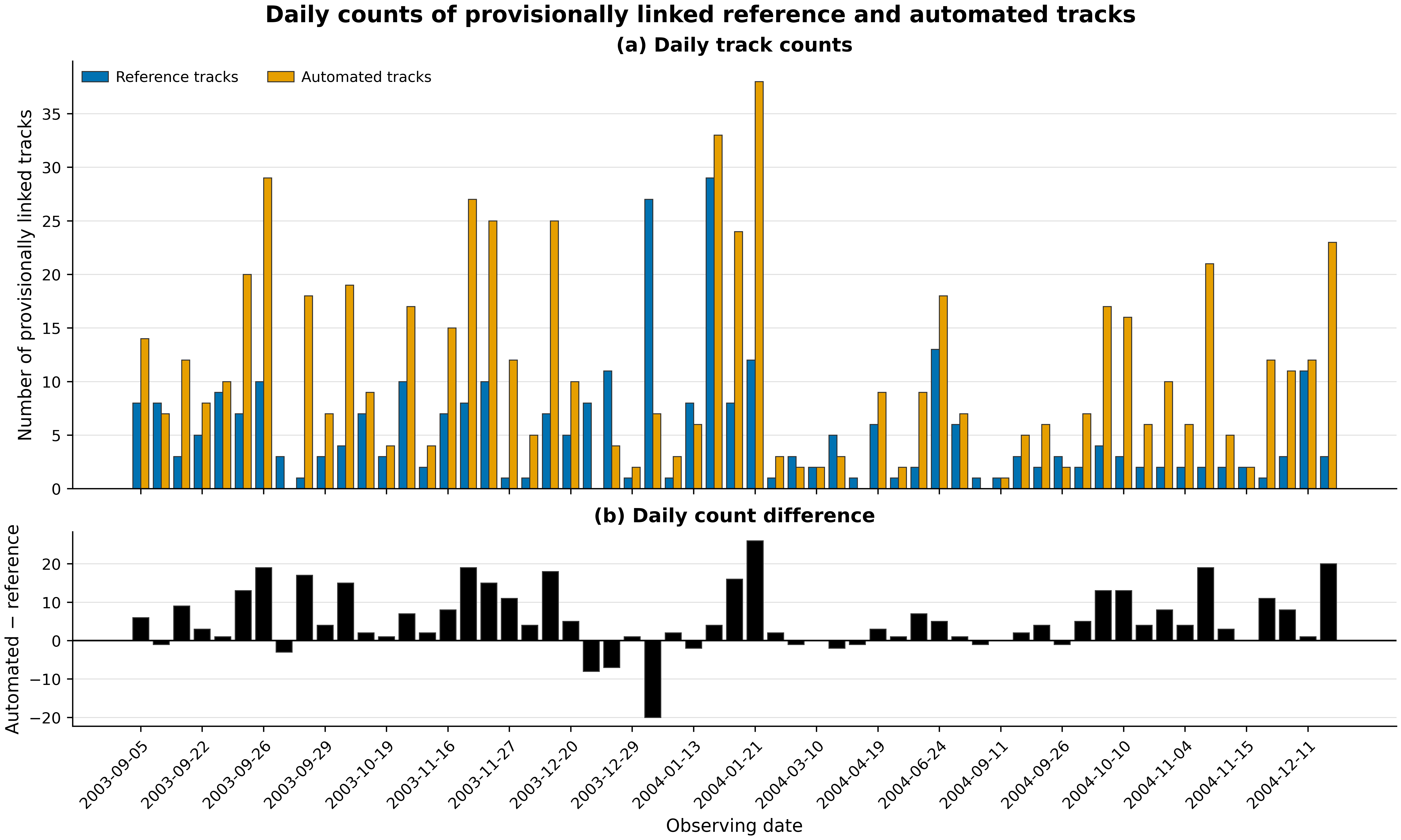}
    \caption{\replaced{Daily counts of provisionally linked reference and automated tracks within the selected manually positive validation cohort. The upper panel shows reference-track counts in blue and automated-track counts in orange. The lower panel shows the automated-minus-reference difference as black bars; the horizontal line marks zero difference. Tracks were generated using the 125-pixel-radius region of interest, a model-score threshold greater than 0.99, a maximum temporal separation of 121.5~s, and a maximum centroid displacement of 10 pixels. These counts are conditional on the selected cohort and are not exposure-normalized archive-wide occurrence rates.}{Daily ripple counts and model–human detection differences over time (2003–2004). Teal circles represent daily ripple counts predicted by the machine learning model, red crosses denote human-annotated counts, and the black line shows the daily difference (Model - Human) referenced to the right axis. The dashed horizontal line indicates zero difference.}}
    \label{fig:daily_counts}
\end{figure}
\deleted{Event-level validation provides a physically meaningful assessment. The manual catalog contains approximately 720 ripple events over the two-year analysis period. The SE-CNN identified approximately 952 events, of which roughly 650 overlapped spatially and temporally with manual detections. The automated catalog contains 32\% more events than the manual catalog (952 vs 720 events), yielding a relative detection bias of +32\%. This difference likely reflects enhanced sensitivity to low-amplitude ripple signatures rather than systematic over-detection. Visual inspection confirms that many additional model detections correspond to weak but coherent banded structures that may fall near subjective manual thresholds.}
\added{Under the 125-pixel-radius same-frame validation procedure and a model-score threshold greater than 0.99, 469 manually selected positive frames were considered. Of these, 439 contained at least one eligible reference support within the evaluation region, and 415 had a valid approximately two-minute predecessor image. At least one retained automated component overlapped an eligible reference support in 299 of the 415 evaluable frames, corresponding to a frame-level recovery fraction of 72.0\% (Wilson 95\% confidence interval: 67.5\%--76.1\%). At the object level, maximum-cardinality one-to-one matching assigned automated components to 324 of 559 eligible reference supports, corresponding to a manual-object recall of 58.0\% (Wilson 95\% confidence interval: 53.8\%--62.0\%). The procedure retained 871 automated components, of which 324 received one-to-one reference assignments and 547 did not. The automated-component assignment fraction was 37.2\%, but this value is not a precision estimate because the reference catalog is not exhaustive.}

\added{The blinded audit examined all 720 strict-zero-overlap disagreement patches. Among the 177 reference-marked/component-unmatched crops, 82 were classified as clear ripple-like and 95 as non-ripple or artifact. The clear-ripple fraction was 46.3\%, with a night-cluster bootstrap 95\% interval of 35.3\%--57.3\%.}

\added{Among the 543 model-candidate/reference-unmarked crops, 411 were classified as clear ripple-like and 132 as non-ripple or artifact. The clear-ripple fraction was therefore 411 of 543, or 75.7\%, with a night-cluster bootstrap 95\% interval of 70.4\%--80.2\%. The clear-ripple fraction was descriptively 29.4 percentage points higher in the model-candidate/reference-unmarked stratum than in the reference-marked/component-unmatched stratum; the bootstrap interval for the difference was 16.6--41.0 percentage points. Because the strata were selected for opposite disagreement outcomes and used different crop-centering rules, this difference is descriptive and is not a controlled comparison of reference and automated performance.}

\begin{table}[htbp]
\centering
\caption{\added{Primary morphology labels from the single-reviewer blinded audit of strict-zero-overlap disagreement patches.}}
\label{tab:disagreement_audit}
\small
\begin{tabular}{lrrrrl}
\toprule
\added{Audit stratum} & \added{Total} & \added{Clear} & \added{Non-ripple/artifact} & \added{Clear-ripple fraction} \\
\midrule
\added{Reference-marked/component-unmatched} & \added{177} & \added{82} & \added{95} & \added{46.3\% (35.3\%--57.3\%)} \\
\added{Model-candidate/reference-unmarked} & \added{543} & \added{411} & \added{132} & \added{75.7\% (70.4\%--80.2\%)} \\
\bottomrule
\end{tabular}
\end{table}

\begin{figure}
    \centering
    \includegraphics[width=\linewidth]{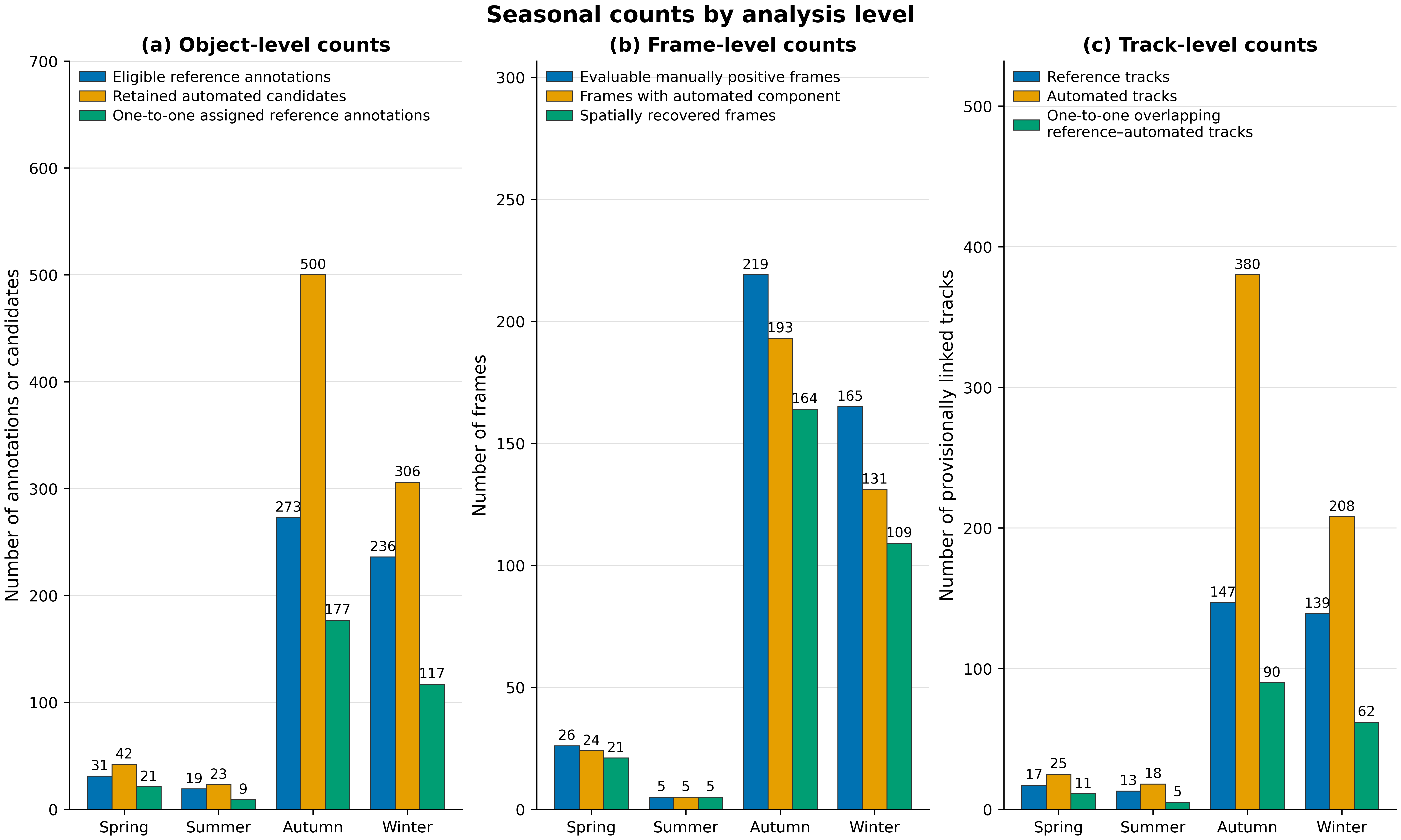}
    \caption{\replaced{Seasonal comparison of object-level, frame-level, and provisionally linked track counts within the selected manually positive validation cohort using a centered 125-pixel-radius region of interest. (a) Counts of eligible reference annotations, retained automated candidates, and reference annotations receiving one-to-one automated assignments. Automated candidates satisfy a model-score threshold greater than 0.99, a pre-clipping area of at least 125 pixels, and an eccentricity no greater than 0.95. (b) Counts of evaluable manually positive frames, frames containing at least one retained automated component, and spatially recovered frames containing at least one reference--automated overlap. (c) Counts of reference tracks, automated tracks, and one-to-one overlapping reference--automated tracks. Provisional tracks were constructed within an observing night by linking detections in consecutive frames when centroid displacement was no greater than 10 pixels and temporal separation was no greater than 121.5~s. These results characterize the selected validation cohort and should not be interpreted as full-archive seasonal occurrence rates.}{Seasonal comparison of ripple-object, frame-level, and event-level counts within the manually selected positive validation cohort using a 125-pixel-radius region of interest. (a) Counts of eligible manual ripple annotations, retained model candidates, and model-recognized manual ripples. Model candidates satisfy a score threshold of \(>0.99\), a pre-clipping area of \(\geq 100\) pixels, and an eccentricity of \(\leq 0.95\). A manual ripple is model-recognized when its \(41\times41\)-pixel support overlaps a retained model component by at least one pixel under one-to-one object assignment. (b) Counts of evaluable manual-positive frames, frames containing at least one retained model component, and spatially recovered frames containing at least one manual--model overlap. (c) Counts of manual-positive events, model-positive events, and overlapping manual--model events. Unique event tracks are constructed within the same observing night by linking detections in consecutive frames when their centroid displacement is \(\leq 10\) pixels and their temporal separation is \(\leq 121.5\)~s. Overlapping events are assigned using maximum-cardinality one-to-one matching between manual and model tracks that share at least one spatially overlapping member detection. Numbers above the bars give the corresponding counts. These results characterize the composition and conditional recovery of the selected validation cohort and should not be interpreted as full-archive seasonal ripple occurrence rates.}}
    \label{fig:ripple_counts}
\end{figure}

\begin{figure}
    \centering
    \includegraphics[width=\linewidth]{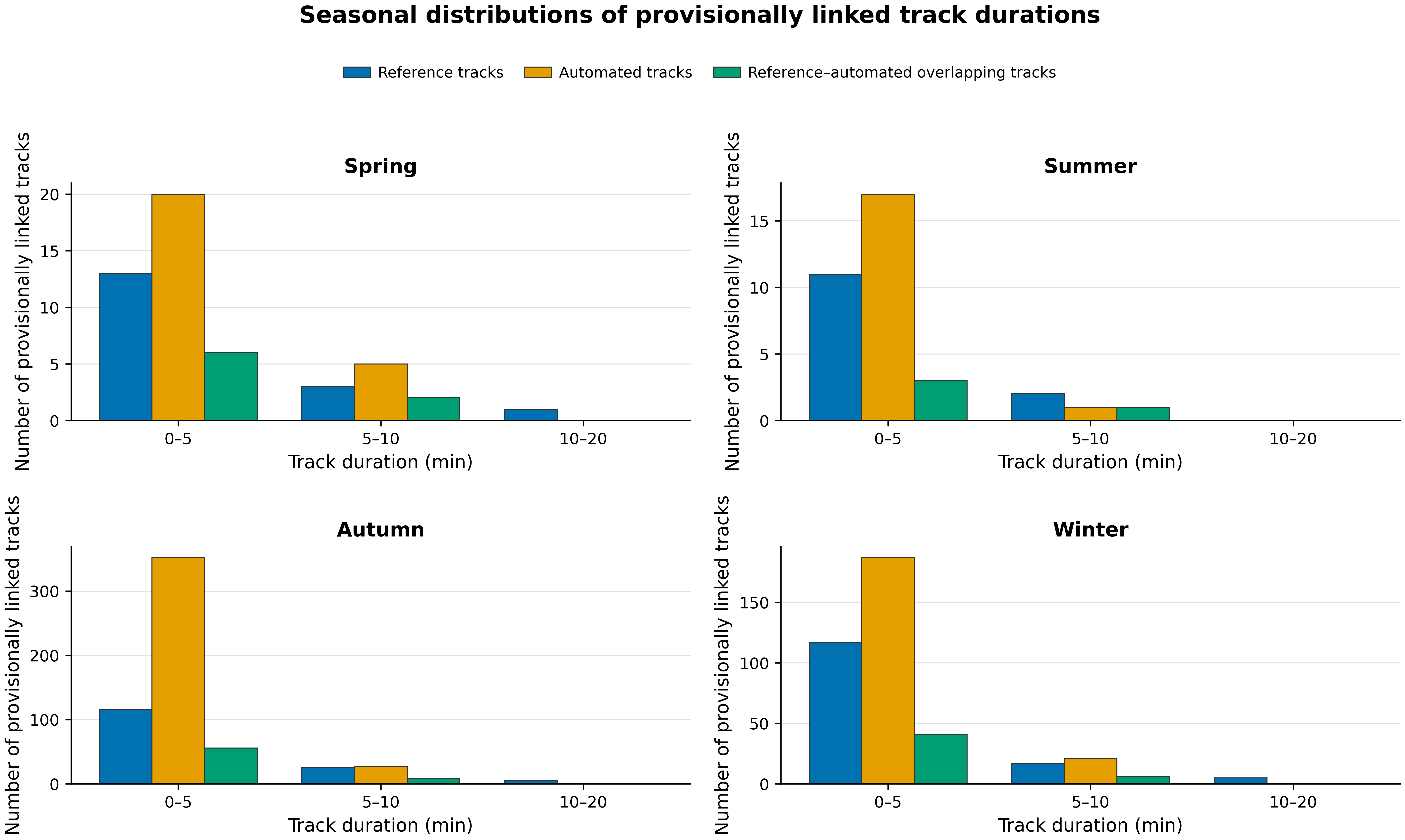}
    \caption{Seasonal distributions of provisionally linked track durations for reference tracks, automated tracks, and one-to-one reference–
  automated overlapping track pairs whose durations fall within the same displayed bin. Track durations were produced by the provisional
  temporal-linking procedure and are not independently verified physical event lifetimes. Results are conditional on the selected manually
  positive validation cohort; counts are descriptive and are not exposure-normalized seasonal occurrence rates.}
    \label{fig:ripple_life_time}
\end{figure}
\deleted{Temporal variability is reproduced with high fidelity. Daily ripple counts derived from the SE-CNN and the manual catalog are strongly correlated (Pearson $r = 0.84$, $p < 0.001$), demonstrating that the automated framework accurately captures both active and quiescent intervals. High-activity periods during autumn 2003 and winter 2004 are consistently identified in both datasets, although peak amplitudes are larger in the automated counts due to enhanced sensitivity.
}
\added{Within the selected validation cohort, daily counts from the provisional reference and automated track catalogs were positively correlated (Pearson $r=0.84$, $p<0.001$; Figure~\ref{fig:daily_counts}). Both catalogs contained elevated counts during portions of autumn 2003 and winter 2004, although automated counts were frequently larger. Because the analysis was restricted to selected manually positive observations, this correlation does not establish performance during quiet archive-wide conditions or identify the cause of the count differences.}

\deleted{Seasonal statistics show consistent large-scale structure. Both methods identify autumn as the dominant season for ripple occurrence, with approximately 610 model detections compared to 360 manual detections. Winter activity is comparable between methods (270 model vs 280 human). Spring activity is modest (50 vs 45), while summer exhibits the lowest occurrence (22 model vs 35 human). The relative seasonal ranking ($Autumn > Winter > Spring > Summer$) is preserved, confirming that the automated framework reproduces the climatological modulation reported by \cite{li2017characteristics}.}
\added{Seasons were defined as spring (March--May), summer (June--August), autumn (September--November), and winter (December--February). Seasonal counts of provisionally linked reference and automated tracks within the selected manually positive validation cohort are shown in Figure~\ref{fig:ripple_counts} and Table~\ref{tab:seasonal_events}. Under the 125-pixel-radius, model-score-greater-than-0.99 configuration and the provisional 10-pixel, 121.5-s linking rule, autumn contained 380 automated tracks and 147 reference tracks. The difference of 233 tracks represented 74.0\% of the summed positive automated-minus-reference differences across seasons. The automated-to-reference count ratio was 2.59 in autumn, compared with 1.47 in spring, 1.38 in summer, and 1.50 in winter. These values are conditional track counts rather than exposure-normalized occurrence rates because the analyzed frames were selected from manually positive observations rather than complete observing nights.}

\added{The patch audit was analyzed separately from the provisional track counts. Among 320 autumn model-candidate/reference-unmarked patches, 243 were classified as clear ripple-like and 77 as non-ripple or artifact, corresponding to a clear-ripple fraction of 75.9\%. The patch audit did not determine whether multiple crops represented duplicate or fragmented temporal tracks.}

\begin{table}[htbp]
  \centering
  \caption{\replaced{Seasonal counts of provisionally linked reference and automated
  tracks}{Seasonal counts of provisionally linked manual and automated
  event tracks} in the selected manually positive validation cohort.
  Tracks were constructed using the radius-125 detector, a score threshold
  greater than 0.99, a maximum temporal gap of 121.5~s, and a maximum
  centroid displacement of 10 pixels. These conditional counts are not
  exposure-normalized seasonal occurrence rates.}
  \label{tab:seasonal_events}
  \begin{tabular}{lrrrrr}
  \toprule
  Season &
  \replaced{Reference}{Manual} &
  Automated &
  Difference &
  \replaced{Automated/reference}{Automated/manual} &
  \replaced{Share of positive difference (\%)}{Excess (\%)} \\
  \midrule
  Spring & 17  & 25  & 8   & 1.47 & 2.5 \\
  Summer & 13  & 18  & 5   & 1.38 & 1.6 \\
  Autumn & 147 & 380 & 233 & 2.59 & 74.0 \\
  Winter & 139 & 208 & 69  & 1.50 & 21.9 \\
  \midrule
  Total  & 316 & 631 & 315 & 2.00 & 100.0 \\
  \bottomrule
  \end{tabular}

  \begin{minipage}{0.95\linewidth}
  \footnotesize
  \textit{Note.} \replaced{Difference is the automated count minus the reference count. Share of positive difference (\%) is the seasonal difference divided by the sum of the positive seasonal differences.}{Difference is the automated count minus the manual count. Excess (\%) is the seasonal difference divided by the sum of the positive seasonal differences.} Track identities are provisionally derived from the temporal-linking algorithm rather than from independently curated physical events.
  \end{minipage}
  \end{table}

\begin{table}[htbp]
\centering
\caption{\added{Seasonal primary-label results for model-candidate/reference-unmarked patches in the single-reviewer blinded audit. Percentages in the clear ripple-like and non-ripple/artifact columns use all candidates in the corresponding season.}}
\label{tab:seasonal_patch_audit}
\begin{tabular}{lrrrrr}
\toprule
\added{Season} & \added{Candidates} & \added{Clear ripple-like} & \added{Non-ripple/artifact} & \added{Clear (\%)} \\
\midrule
\added{Spring} & \added{21} & \added{11 (52.4\%)} & \added{10 (47.6\%)} & \added{52.4} \\
\added{Summer} & \added{14} & \added{9 (64.3\%)} & \added{5 (35.7\%)} & \added{61.5} \\
\added{Autumn} & \added{320} & \added{243 (75.9\%)} & \added{77 (24.1\%)} & \added{75.6} \\
\added{Winter} & \added{188} & \added{148 (78.7\%)} & \added{40 (21.3\%)} & \added{78.6} \\
\midrule
\added{Total} & \added{543} & \added{411 (75.7\%)}& \added{132 (24.3\%)} & \added{75.7} \\
\bottomrule
\end{tabular}
\begin{minipage}{0.95\linewidth}
\footnotesize
\added{\textit{Note.} The audit concerns selected $41\times41$-pixel patches rather than linked tracks or an archive-wide sample. Results reflect one reviewer and are not consensus-adjudicated labels. Duplicate and fragmented track representations were not assessed.}
\end{minipage}
\end{table}

%

\deleted{Lifetime distributions further demonstrate consistency. Indicated by Figure 7, short-lived events (0–5 min) dominate both catalogs (466 model vs 362 human) in Autumn, and the overall distribution shape is similar across seasons. The automated method identifies a slightly larger fraction of very short-lived ripples, whereas the manual catalog includes a marginally higher fraction of longer-duration events in winter. These differences likely arise from the patch-based detection framework, which emphasizes localized spatial texture and may fragment extended events into shorter segments.}

\added{Figure \ref{fig:ripple_life_time} shows the seasonal distributions of durations produced by the provisional temporal-linking procedure. Most tracks occurred in the
0–5 min bin: 257 of 316 reference tracks (81.3\%) and 576 of 631 automated tracks (91.3\%). The 5–10 min bin contained 48 reference and 54 automated tracks, while the 10–20 min bin contained 11 reference tracks and one automated track. The fraction of automated tracks in the 0–5 min bin was higher than the corresponding reference-track fraction in all four seasons. Of the 168 one-to-one overlapping reference–automated track pairs, 124 (73.8\%) had both durations in the same displayed bin and are represented by the green bars; the remaining 44 pairs occupied different duration bins. These counts are conditional on the selected manually positive validation cohort and are not normalized for seasonal observing exposure.}

\deleted{Figure 8 presents the seasonal distributions of ripple propagation directions derived from both the automated detection
model and human annotations. The results reveal clear seasonal anisotropy in ripple motion, indicating that propagation
is not randomly distributed in azimuth but instead modulated by background atmospheric conditions. During autumn
and winter—the seasons with the highest ripple occurrence—the distributions exhibit enhanced meridional components,
with notable peaks in the southward and southwestward sectors. The automated model reproduces these dominant
directional tendencies with good consistency relative to human identification, particularly in autumn when the sample
size is largest. In contrast, summer exhibits a comparatively different directional structure, with weaker meridional
dominance and a more pronounced zonal component in the human-identified events. The model captures the general
variability but shows some discrepancies in the relative amplitude of specific sectors during low-activity seasons.
Spring represents a transitional regime, characterized by broader directional spread and reduced coherence. Overall,
the seasonal dependence of ripple propagation direction suggests that mesospheric instability processes are strongly
influenced by seasonal changes in background winds and gravity-wave propagations.}

\deleted{Overall, across independent metrics — event recovery rate (90\%), strong daily correlation (r = 0.84), consistent seasonal ranking, and comparable lifetime structure — the SE-CNN framework faithfully reproduces the ripple characteristics
reported by \cite{li2017characteristics}. The principal difference is increased sensitivity to faint, short-lived instability signatures, resulting in a 32\% higher event count. These results demonstrate that the automated detection system is not only consistent with established manual analysis but also provides a more objective and scalable tool for constructing long-term climatology of mesospheric instability activity.}

\section{\replaced{Discussion and Summary}{Discussion}}
\deleted{We trained a CNN classifier coupled with squeeze-and-excitation block on the labeled patch dataset. The squeeze-and-excitation (SE) block was incorporated to enhance the network’s ability to distinguish subtle ripple-scale structures from background variability in airglow images. Ripple patterns are characterized by faint, banded intensity modulations embedded within larger-scale wave structures and nonuniform background emissions. Standard convolutional layers treat all feature channels equally, which may limit sensitivity to weak but physically meaningful spatial signatures. The SE mechanism introduces adaptive channel-wise weighting through global context aggregation, allowing the network to recalibrate feature responses based on their relative importance. In the context of ripple detection, this enables the model to emphasize channels that capture coherent banded structures while suppressing noise-dominated or large-scale background features. By dynamically modulating feature importance, the SE block improves discrimination between true instability-driven ripple patterns and spurious intensity fluctuations, thereby enhancing detection robustness.}
\added{The SE-CNN was designed to identify weak, locally coherent banded morphology in normalized time-difference images. The SE blocks recalibrate learned feature channels using globally aggregated channel descriptors, allowing different weights to be assigned to the extracted feature maps. Figure~\ref{fig:cnn-SE-comparison} compares this architecture with a baseline CNN that omits the SE blocks. The comparison evaluates the effect of channel recalibration under the stated patch-sampling and training procedure. It does not determine whether an individual detected structure is an instability feature, a secondary gravity wave, or another image morphology.}

\deleted{The automated SE-CNN detection framework demonstrates strong quantitative agreement with the manually curated ripple catalog of Li et al. (2017). Event-level recall of approximately 90\% indicates that the vast majority of manually identified ripple events are successfully recovered by the model. Daily ripple counts are highly correlated ($r = 0.84$, $p < 0.001$), confirming that the automated system preserves the temporal structure of ripple activity, including both high-activity and quiescent intervals. The short-dominated lifetime structure also closely match previously reported statistics. These results collectively indicate that the SE-CNN framework faithfully reproduces the principal observational properties established in earlier manual analyses.}
\added{The validation results characterize agreement at several distinct analysis levels. On the held-out patch test set, the SE-CNN achieved an F1 score of 0.92 under the stated positive- and background-patch sampling procedure. Within the selected manually positive validation cohort, 72.0\% of evaluable frames contained at least one reference-automated spatial overlap, and 58.0\% of eligible reference supports received one-to-one automated-component assignments. The disagreement audit further showed that many high-score automated candidates without a reference mark contained clear ripple-like morphology in the reviewed crop. These measures describe different analysis units and should not be combined into a single recovery statistic.}

\deleted{}
\added{The present classifier identifies ripple-like morphology in OH airglow images rather than the physical generation mechanism. Although such structures have traditionally been interpreted as signatures of gravity-wave breaking, \citet{li2017characteristics} showed that similar small-scale patterns may also represent secondary gravity waves when additional wind and temperature measurements are considered. Because the present analysis uses only airglow imagery, the automated labels are morphological rather than physical classifications.}

\deleted{The observed climatology of ripple occurrence reflects seasonal modulation of mesospheric dynamical instability by background winds, gravity wave activity, and tidal forcing. The pronounced autumn maximum and enhanced winter activity suggest that ripple formation is favored during periods of stronger gravity wave filtering and increased wind shear in the mesopause region. Ripple occurrence peaks in autumn when tidal amplitudes and gravity wave perturbations are relatively strong, enhancing the likelihood of dynamic instability (low Richardson number) or convective instability (negative $N^2$). The reduced occurrence during summer indicates more stable background conditions or less favorable shear configurations for wave breaking.}
\added{Within the selected manually positive validation cohort, the largest automated--reference track-count difference occurred during autumn. Autumn also contained the largest number of retained candidates, which may amplify the absolute difference. The seasonal counts depend on the cohort composition, model-score threshold, region of interest, and provisional temporal-linking criteria, and they are not normalized by usable observing exposure. They therefore do not establish a unique physical cause or a full-archive climatology. Possible methodological contributors include seasonal differences in background and image quality, training-data composition, reference-annotation practice, and fragmentation or duplication during linking. Without coincident wind and temperature measurements, the autumn difference cannot be assigned uniquely to tides, background winds, gravity-wave filtering, instability, or secondary-wave generation.}

\added{The duration distributions should be interpreted as properties of the provisional tracking algorithm rather than measurements of physical event lifetimes. In particular, the larger fraction of automated tracks in the 0–5 min bin may reflect differences in detection continuity, temporary score-threshold crossings, fragmentation of automated components, or the temporal-linking rule. Conversely, longer linked tracks could result from repeated association of nearby detections and do not by themselves establish continuous physical persistence. The difference between reference and automated duration-bin assignments among some one-to-one overlapping pairs further indicates that spatial overlap does not guarantee identical temporal extent. Establishing physical event durations would require independent temporal adjudication of track identity and continuity.}

\added{Previous studies show that the reported occurrence and morphology of ripple-like structures vary among sites, instruments, and analysis methods. \citet{yue2010seasonal} reported seasonal and local-time variability in multi-year OH airglow observations from Colorado and Japan, while \citet{suzuki2011short} and \citet{rourke2017climatological} found different temporal and directional behavior at Antarctic sites. Machine-learning studies also use different targets: \citet{lai2019automatic} extracted wave patterns, whereas \citet{prasad2025ripple} and \citet{wust2026extraction} used object-detection architectures to localize confined wave-like structures. Their reported precision, recall, occurrence measures, and seasonal distributions are not directly interchangeable with the present results because the instruments, labels, preprocessing, thresholds, occurrence denominators, and evaluation units differ. Unlike the object-detection approaches of \citet{prasad2025ripple, wust2026extraction}, the present framework first classifies fixed-size image patches and then reconstructs candidate regions through overlapping sliding-window inference.}

\deleted{Importantly, the automated approach offers methodological advantages beyond simple replication of manual results. First, it removes observer-dependent variability by applying a consistent decision boundary across the entire dataset. Second, it enhances sensitivity to low-amplitude ripple signatures, potentially revealing instability activity that is underrepresented in manually curated catalogs. Third, the computational efficiency of the SE-CNN framework enables rapid processing of extended multi-year archives, making construction of long-term ripple climatologies feasible. Such scalability is essential for investigating interannual variability, coupling with meteor radar wind measurements, and potential climate-scale trends in mesospheric instability occurrence.}
\added{The principal methodological advantage of the automated approach is consistent and scalable application of a learned morphological criterion. Because the classifier was trained on manually labeled examples, it inherits the definitions and uncertainties of those labels. Automation nevertheless avoids repeatedly reapplying the criterion through separate manual inspections and enables reproducible processing of large image collections. Archive-wide occurrence rates and long-term climatological analyses would require representative sampling of both positive and quiet frames, exposure normalization, and validation beyond the manually positive cohort considered here.}

\deleted{Several limitations warrant consideration. The classifier is trained on manually labeled examples from a specific site, and its performance may depend on the representativeness of the training data. Imaging conditions such as auroral contamination, cloud interference, or extremely faint ripple structures may reduce detection reliability. In addition, the patch-based detection and clustering approach provides event occurrence and approximate location but does not directly yield high-precision geometric measurements of ripple amplitude or wavelength. Future work could integrate object-detection or segmentation frameworks to improve spatial delineation of ripple fronts and enable automated extraction of quantitative wave parameters.}
\added{Several limitations should be considered. The classifier was trained using manually labeled examples from one site, and its performance depends on the representativeness and accuracy of those labels. The full-image validation cohort contained manually positive frames rather than a prevalence-representative archive sample; archive-wide precision, specificity, and quiet-frame false-alarm rate therefore cannot be estimated. The disagreement audit used one reviewer, so inter-reviewer agreement and consensus labels are unavailable, and some $41\times41$-pixel crops may require wider spatial or temporal context. The reference-marked/component-unmatched stratum was defined by overlap with a final retained component rather than by the classifier score at the exact reference coordinate. Finally, the temporal tracks are products of the provisional linking algorithm rather than independently verified continuous physical events. The method identifies image morphology and does not determine the underlying generation mechanism.}

\deleted{Overall, the strong agreement across multiple independent validation metrics — including event recovery rate, temporal correlation and seasonal modulation — demonstrates that the SE-CNN framework provides a reliable and objective surrogate for manual ripple identification. The enhanced sensitivity and scalability of the automated approach establish a robust foundation for comprehensive, long-term studies of mesospheric gravity wave instability dynamics.}
\added{The SE-CNN provides a reproducible implementation of a manually defined morphological criterion for detecting ripple-like structures in OH airglow images. Validation at the held-out patch, frame, object, and disagreement-patch levels demonstrates substantial but incomplete agreement with the reference catalog. The audit further shows that disagreement has multiple possible sources, including additional model-detected morphology, reference-label or coordinate uncertainty, limited crop context, thresholding, and component filtering. Broader deployment will require representative archive-wide sampling, quiet-frame evaluation, multi-reviewer adjudication, and external validation at additional observing sites.}

\textit{Code and data availability} 
\added{The reconstructed patch dataset contains 725 reference-centered positive patches and 725 deterministically sampled background patches before augmentation. The exact training/validation/test split manifest, preprocessing code, trained model weights, and analysis scripts are archived at DOI~10.5281/zenodo.18927628 and at \url{https://github.com/Multi-Scale-Wave-Dynamics-Group/automated-ripple-recognition.git}. The archived audit materials include the completed primary-label table, blind-ID manifest, review instructions, label-validation script, night-cluster bootstrap script, random seed, and final analysis configuration. The SHA-256 checksum of the completed audit label file is \texttt{26f51078c808a917bc04ff774afc93d982b91d3347a68d0e02916e2afb0908ff}. Secondary interface fields that remained at automatic preset values were not analyzed as reviewer judgments. Raw YRFS OH airglow images are available from the data provider subject to the applicable data-use conditions.}

\textit{Author contribution} \\
Conceptualization: A. Z. Liu \\
Methodology: J. Hu, A. Z. Liu, and W. Dong \\
Software: J. Hu and A. Z. Liu \\
Validation: J. Hu, A. Z. Liu, and W. Dong \\ 
Formal analysis: J. Hu\\
Investigation: J. Hu \\
Resources: A. Z. Liu\\ 
Data curation: A. Z. Liu, A. Feener, J. Li, and T. Li\\ 
Writing -- original draft: J. Hu\\ 
Writing -- review \& editing: J. Hu, A. Z. Liu, and W. Dong\\
Visualization: J. Hu\\ 
Supervision: A. Z. Liu and W. Dong\\ 
Project administration: J. Hu\\ 
Funding acquisition: W. Dong\\

\textit{Competing interests} The authors declare that they have no competing interests.

\textit{Acknowledgments} The project is funded by NASA Grant 80NSSC24K0124 and NSF Grant AGS-2327914. 
\bibliographystyle{copernicus}
\bibliography{agusample}

@article{baker1988rocket,
  title={Rocket measurements of the altitude distributions of the hydroxyl airglow},
  author={Baker, Doran J and Stair Jr, AT},
  journal={Physica scripta},
  volume={37},
  number={4},
  pages={611--622},
  year={1988}
}

@article{fritts2003gravity,
  title={Gravity wave dynamics and effects in the middle atmosphere},
  author={Fritts, David C and Alexander, M Joan},
  journal={Reviews of geophysics},
  volume={41},
  number={1},
  year={2003},
  publisher={Wiley Online Library}
}

@inproceedings{garcia2017mesospheric,
  title={Mesospheric OH layer altitude at midlatitudes: variability over the Sierra Nevada Observatory in Granada, Spain},
  author={Garc{\'\i}a-Comas, Maya and L{\'o}pez-Gonz{\'a}lez, Mar{\'\i}a Jos{\'e} and Gonz{\'a}lez-Galindo, Francisco and de la Rosa, Jos{\'e} Luis and L{\'o}pez-Puertas, Manuel and Shepherd, Marianna G and Shepherd, Gordon G},
  booktitle={Annales Geophysicae},
  volume={35},
  number={5},
  pages={1151--1164},
  year={2017},
  organization={Copernicus Publications G{\"o}ttingen, Germany}
}

@inproceedings{hu2018squeeze,
  title={Squeeze-and-excitation networks},
  author={Hu, Jie and Shen, Li and Sun, Gang},
  booktitle={Proceedings of the IEEE conference on computer vision and pattern recognition},
  pages={7132--7141},
  year={2018}
}

@article{li2017characteristics,
  title={Characteristics of ripple structures revealed in OH airglow images},
  author={Li, Jing and Li, Tao and Dou, Xiankang and Fang, Xin and Cao, Bing and She, Chiao-Yao and Nakamura, Takuji and Manson, Alan and Meek, Chris and Thorsen, Denise},
  journal={Journal of Geophysical Research: Space Physics},
  volume={122},
  number={3},
  pages={3748--3759},
  year={2017},
  publisher={Wiley Online Library}
}

@article{li2005concurrent,
  title={Concurrent OH imager and sodium temperature/wind lidar observation of localized ripples over northern Colorado},
  author={Li, Tao and She, CY and Williams, Bifford P and Yuan, Tao and Collins, Richard L and Kieffaber, Lois M and Peterson, Alan W},
  journal={Journal of Geophysical Research: Atmospheres},
  volume={110},
  number={D13},
  year={2005},
  publisher={Wiley Online Library}
}

@article{lai2019automatic,
  title={Automatic extraction of gravity waves from all-sky airglow image based on machine learning},
  author={Lai, Chang and Xu, Jiyao and Yue, Jia and Yuan, Wei and Liu, Xiao and Li, Wei and Li, Qinzeng},
  journal={Remote Sensing},
  volume={11},
  number={13},
  pages={1516},
  year={2019},
  publisher={MDPI}
}

@article{nakamura2005simultaneous,
  title={Simultaneous observation of dual-site airglow imagers and a sodium temperature-wind lidar, and effect of atmospheric stability on the airglow structure},
  author={Nakamura, T and Fukushima, T and Tsuda, T and She, C-Y and Williams, BP and Krueger, D and Lyons, W},
  journal={Advances in Space Research},
  volume={35},
  number={11},
  pages={1957--1963},
  year={2005},
  publisher={Elsevier}
}

@article{okui2025cnn,
  author = {Okui, Haruka and Wright, Corwin J. and Berthelemy, Peter G. and Hindley, Neil P. and Hoffmann, Lars and Barnes, Andrew P.},
  title = {A Convolutional Neural Network for the Detection of Gravity Waves in Satellite Observations and Numerical Simulations},
  journal = {Geophysical Research Letters},
  year = {2025},
  volume = {52},
  pages = {e2025GL115683}
}

@inproceedings{prasad2025ripple,
  title={Ripple Identification in OH Airglow Images Using YOLO},
  author={Prasad, Ritesh and Bandyopadhyay, Oishila and Singh, Ravindra Pratap and Das, Uma},
  booktitle={2025 IEEE International Conference on Next-Gen Technologies of Artificial Intelligence and Geoscience Remote Sensing (EarthSense)},
  pages={1--5},
  year={2025},
  organization={IEEE}
}

@article{rourke2017climatological,
  title={A climatological study of short-period gravity waves and ripples at Davis Station, Antarctica (68 S, 78 E), during the (austral winter February--October) period 1999--2013},
  author={Rourke, S and Mulligan, FJ and French, WJR and Murphy, DJ},
  journal={Journal of Geophysical Research: Atmospheres},
  volume={122},
  number={21},
  pages={11--388},
  year={2017},
  publisher={Wiley Online Library}
}

@article{suzuki2011short,
  title={Short-period gravity waves and ripples in the South Pole mesosphere},
  author={Suzuki, S and Tsutsumi, M and Palo, SE and Ebihara, Y and Taguchi, M and Ejiri, M},
  journal={Journal of Geophysical Research: Atmospheres},
  volume={116},
  number={D19},
  year={2011},
  publisher={Wiley Online Library}
}

@article{shi2019diurnal,
  title={Diurnal and nocturnal cloud segmentation of all-sky imager (ASI) images using enhancement fully convolutional networks},
  author={Shi, Chaojun and Zhou, Yatong and Qiu, Bo and He, Jingfei and Ding, Mu and Wei, Shiya},
  journal={Atmospheric Measurement Techniques},
  volume={12},
  number={9},
  pages={4713--4724},
  year={2019},
  publisher={Copernicus Publications G{\"o}ttingen, Germany}
}

@article{taylor1990origin,
  title={On the origin of ripple-type wave structure in the OH nightglow emission},
  author={Taylor, Michael J and Hapgood, MA},
  journal={Planetary and space science},
  volume={38},
  number={11},
  pages={1421--1430},
  year={1990},
  publisher={Elsevier}
}

@article{taylor1991near,
  title={Near infrared imaging of hydroxyl wave structure over an ocean site at low latitudes},
  author={Taylor, Michael J and Hill, MJ},
  journal={Geophysical Research Letters},
  volume={18},
  number={7},
  pages={1333--1336},
  year={1991},
  publisher={Wiley Online Library}
}

@article{von2015lunar,
  title={Lunar semidiurnal tide in the terrestrial airglow},
  author={von Savigny, C and Lednyts' kyy, O and Forbes, JM and Zhang, X},
  journal={Geophysical Research Letters},
  volume={42},
  number={9},
  pages={3553--3559},
  year={2015},
  publisher={Wiley Online Library}
}

@article{wust2020variability,
  title={Variability of the Brunt--V{\"a}is{\"a}l{\"a} frequency at the OH-airglow layer height at low and midlatitudes},
  author={W{\"u}st, Sabine and Bittner, Michael and Yee, Jeng-Hwa and Mlynczak, Martin G and Russell III, James M},
  journal={Atmospheric Measurement Techniques},
  volume={13},
  number={11},
  pages={6067--6093},
  year={2020},
  publisher={Copernicus GmbH}
}

@article{wust2026extraction,
  title={Extraction of spatially confined small-scale waves from high-resolution all-sky airglow images based on machine learning},
  author={W{\"u}st, Sabine and Strutz, Jakob and Hannawald, Patrick and Steffen, Jonas and Lienhart, Rainer and Bittner, Michael},
  journal={Atmospheric Measurement Techniques},
  volume={19},
  number={10},
  pages={3539--3556},
  year={2026},
  publisher={Copernicus Publications G{\"o}ttingen, Germany}
}

@article{xiao2024atmospheric,
  author = {Xiao, Beimin and Hu, Shensen and Ai, Weihua and Li, Yi},
  title = {Atmospheric Gravity Wave Detection in Low-Light Images: A Transfer Learning Approach},
  journal = {Electronics},
  year = {2024},
  volume = {13},
  pages = {4030}
}

@inproceedings{yue2010seasonal,
  title={Seasonal and local time variability of ripples from airglow imager observations in US and Japan},
  author={Yue, Jia and Nakamura, Takuij and She, C-Y and Weber, Maria and Lyons, Walter and Li, Tao},
  booktitle={Annales Geophysicae},
  volume={28},
  number={7},
  pages={1401--1408},
  year={2010},
  organization={Copernicus Publications G{\"o}ttingen, Germany}
}

@article{yamada2001breaking,
  title={Breaking of small-scale gravity wave and transition to turbulence observed in OH airglow},
  author={Yamada, Y and Fukunishi, H and Nakamura, T and Tsuda, T},
  journal={Geophysical research letters},
  volume={28},
  number={11},
  pages={2153--2156},
  year={2001},
  publisher={Wiley Online Library}
}

\end{document}